\documentclass[aps,superscriptaddress,twocolumn]{revtex4}
\usepackage{hyperref,amssymb,graphicx,bbm}

\usepackage{xcolor}

\begin{document}
\title{A new derivation of the Hénon's isochrone potentials}

\author{Alberto Saa}\email{asaa@ime.unicamp.br}
\address{Department of Applied Mathematics, University of Campinas, 13083-859 Campinas, SP, Brazil.}
\author{Roberto Venegeroles}\email{roberto.venegeroles@ufabc.edu.br}
\address{Centro de Matem\'atica, Computa\c c\~ao e Cogni\c c\~ao, Universidade Federal do ABC, 09210-580, Santo Andr\'e, SP, Brazil}

\date{\today}

\begin{abstract}
We revisit in this note the Hénon’s isochrone problem. By using the standard Abel inversion technique for one-dimensional motion, we 
{recover in a simple way the Hénon's parabolae and}
 get all isochrone central potentials  under mild smoothness assumptions on the potential function.  
Our approach also allows us to conclude that isochronous radial periods with explicit energy dependence   are necessarily Keplerian, {\em i.e.}, $T^{2}\propto|E|^{-3}$, and that their corresponding orbits can be easily integrated  by mapping them into the usual Kepler problem. It can also be employed to study some other inverse central-force problems and, in
particular, it  provides a     proof of  Bertrand's theorem.
\end{abstract}


\maketitle

\section{Introduction}
\label{sec1}

The concept of   isochrone potentials   was put forward by Michel H\'enon in the fifties  \cite{MH1,MH2,MH3} in the study of globular clusters,  see \cite{JBY} for a brief review of the subject. As a first approximation, one models dynamics within these dense, approximately spherically symmetric, conglomerations of stars  by means of an averaged potential acting on any one of the stars due to the influence of all the others, leading naturally in this way  to the study of general   {central  potentials}. Hénon, apparently out of pure mathematical curiosity, investigated what he dubbed as the {\em isochrone problem}: which  {central   potentials} would lead to radial {oscillation periods} that depend only on energy $E$ and not on angular momentum $\ell$? He solved this problem and found  that these {\em isochrone potentials} he derived yielded better models for the averaged potential of globular clusters than those that were dominant at the time. A dynamical mechanism called resonant relaxation was   proposed to explain how the mass distribution of a globular cluster could evolve towards an isochrone model   \cite{JBY}. H\'enon's  pioneering work has attracted considerable attention in recent years and his original approach has been increasingly elucidated and deepened, see \cite{JBY,BTGD,PPD,PPP,PPC,RP}.   Although the modern understanding is that  realistic globular clusters do require a more sophisticated description, the isochrone models are  still inspiring and are being actively used  in astrophysical applications \cite{R1,R2,R3}.   {The original Hénon's parametric derivation of the isochrone potentials
 \cite{MH1,MH2,MH3} is quite involved. Due to its importance,  Hénon's results have been 
 rederived in many distinct ways. For instance, Hénon's problem was solved   in \cite{PPD,PPP} using complex analysis, in
 \cite{PPC} using Euclidean geometry, and in \cite{RP}   using the Hamiltonian formulation of the problem and Birkhoff normal forms. We present here another derivation for   Hénon's isochrone potentials  {based on the Abel inversion technique \cite{Abel} and requiring  milder smoothness assumptions than those ones usually assumed in the literature. 
 Furthermore,  our approach based on the Abel integral inversion is
  quite general  and can be used in  other contexts as well.  }

Any central-force problem is known to be integrable and its dynamics is completely determined by the usual conserved quantities per unit mass $E$ and $\ell$, with
\begin{equation}
\label{ene}
E=\frac{1}{2}\dot{r}^{2}+U(r) ,
\end{equation}
where  $U(r)$ stands for the effective central potential
\begin{equation}
\label{eff}
U(r)=V(r)+\frac{\ell^{2}}{2r^{2}},
\end{equation}
with $V(r)$ being the central potential. The so-called {azimuthal angle \cite{PPD}}
\begin{equation}
\label{apsi}
\Theta(E,\ell)=\int_{r_{\rm min}}^{r_{\rm max}}\frac{\ell}{r^{2}}\frac{1}{\sqrt{2[E-U(r)]}}dr,
\end{equation}
which stands for 
the angular variation between the points of smaller (periapsis) and greater (apoapsis) approximation of the center, and the radial period
\begin{equation}
\label{te}
T(E,\ell)=\sqrt{2}\int_{r_{\rm min}}^{r_{\rm max}}\frac{1}{\sqrt{E-U(r)}}dr,
\end{equation}
are the two fundamental quantities in the analysis.
{Notice that the apsidal angle, defined as
the angular variation during one radial period, is twice the azimuthal angle (\ref{apsi}).}
 The isochrony condition is equivalent to requiring  that the radial period $T$ does not depend on the angular momentum $\ell$.
{By   a judicious analysis of the integral (\ref{te}), Hénon showed that
the isochrony condition requires
\begin{equation}
\label{parabola}
(ax + bY)^2 + cx + dY + e =0,
\end{equation}
where  $x=2r^2$ and $Y(x) = xV\left(\sqrt{x/2}\right)$ are known as the 
  Hénon variables, and $a$, $b$, $c$, $d$, and $e$ are constants which can be expressed
  in terms of the parameters of the original dynamics, see \cite{MH1} for further details. As one can see, the isochrone potentials $V$ are determined by  the parabola (\ref{parabola}) in the $(x,Y)$ plane. Such relation of the isochrone potentials with
  a conic section is still rather mysterious, despite the recent deep geometrical analysis
  of \cite{PPC}. Novel derivations of (\ref{parabola}) are especially welcome since they might provide new insights into a better understanding  of this fundamental relation.}
  
 The Newtonian and isotropic harmonic potentials
\begin{equation}
\label{pot}
V_{\rm Ne}(r) = -\frac{k}{r}, \quad
V_{\rm ha}(r) = \frac{k}{2}r^2,
\end{equation}
are the simplest isochrone potentials, as one can see directly from their well known radial periods
\begin{equation}
T_{\rm Ne}=\frac{\pi k}{\sqrt{2|E|^{3}}}, \quad T_{\rm ha}=\frac{\pi}{\sqrt{k}},
\end{equation}
which clearly do not depend on $\ell$. {It is easy to verify that both potentials (\ref{pot}) belong to the}  
 class of parabolae  given by (\ref{parabola}), which includes both the
 straight and laid types  {\cite{PPD}.} Bertrand's theorem assures that these two potentials are the only ones for which all bounded orbits are closed, {or, equivalently,
they are the only ones for which the azimuthal angle (\ref{apsi}) is a rational multiple of $\pi$ independent of
$\ell$. 
} Hénon’s isochrone problem encompasses the Bertrand’s theorem in the sense that the two central potentials leading to closed orbits are trivially isochrone, but there are other isochrone potentials for which generic bounded orbits are not closed, 
 { {\em i.e.}, the azimuthal angle (\ref{apsi}) will be a real function of 
$\ell$.} In summary, the basic isochrone potentials {derived from (\ref{parabola})} are, besides the Bertrand’s cases (\ref{pot}), the so-called H\'enon potential
\begin{equation}
\label{he}
V_{\rm He}(r) = -\frac{k}{b+\sqrt{b^2 + r^2}},  
\end{equation}
and, respectively, the bounded and hollowed potentials 
\begin{equation}
\label{boho}
V_{\rm bo}(r) = \frac{k}{b + \sqrt{b^2 - r^2}},   \quad
V_{\rm ho}(r) = -\frac{k}{r^2}\sqrt{ r^2- b^2 }.
\end{equation}
It is clear that the bounded and hollowed potentials are not defined for all $r$, and that the Newtonian potential arises in the $b=0$ limits of the H\'enon (\ref{he}) or the hollowed (\ref{boho}) potentials. The potentials (\ref{pot}), (\ref{he}), and (\ref{boho}) do not exhaust the set of all isochrone potentials. In fact, if $V(r)$ is an isochrone potential, then $V(r)+\varepsilon + \Lambda/r^2$ will be also isochrone since the extra terms only imply some shifts in the conserved quantities $E$ and $\ell$. These extra terms are called $(\varepsilon,\Lambda)$-gauges in this context. Despite being a rather fundamental result, all known derivations of the isochrone potentials are quite involved, see, for instance,     \cite{PPC}
for further details.

In the present note, which follows all notations and structure of {\cite{PPD}},  {we present a new derivation of the Hénon parabolae and all the isochrone potentials based on a standard technique from the inverse problem for one-dimensional motion \cite{LL,AHC} and an algebraic equation.}
 {Our derivation assumes a mild smoothness assumption 
on the potential function $V(r)$, in fact,  we only require a continuously differentiable $V(r)$, in contrast with the usual derivations in the literature, which typically require an analytic $V(r)$. This is explicitly the case of
Refs. \cite{JBY} and \cite{PPC}, for instance, where the integrand of (\ref{te}) is expanded in a  Maclaurin series giving origin to terms involving the Wallis
integral, or \cite{PPD}, where the ``parabola property'' is proved in Appendix B under an
explicit    analyticity 
assumption.} Our approach also allows to show  that all isochrone potentials
of the family (\ref{he}) and (\ref{boho}) have Keplerian radial period $T^2\propto |E|^{-3}$, and that their orbits  can be integrated easily by mapping them into an usual Kepler problem.

\section{Inverse Central-Force Problem}
\label{sec2}

 {
In order to introduce the   inverse problem approach, {\em i.e.}, to obtain the effective potential $U$ by means of $\Theta$ and $T$, it is useful to   change the integration variable 
from $r$ to $U$, so that $r_1(U) \le r_2 (U)$ are the two
branches of the inverse function to $U(r)$ near a local minimum $U_0 = U(r_0)$, see Fig. \ref{fig1}.
Notice that $r_1' < 0$ and
$r_2'  > 0$ and that, in terms of $U$, the integration in $r$ reads }
\begin{figure}[b]
\begin{center}
\input{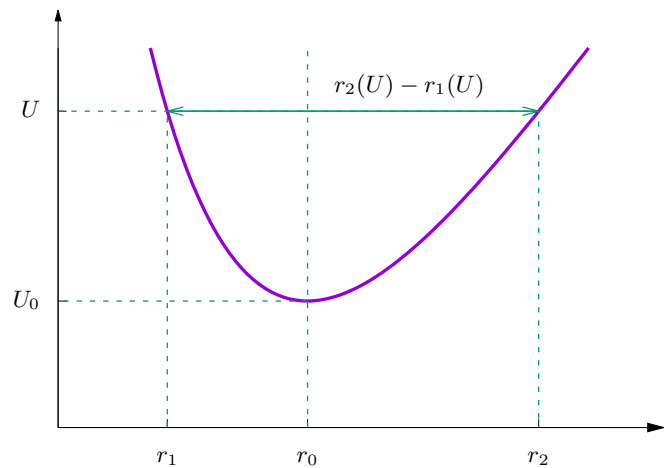}
\end{center}
 \caption{Aspect of a generic effective potential $U(r)$ near its local minimum at $r=r_0$. 
 {The continuous differentiability of $V(r)$ and the existence of the local
minimum of $U(r)$, which corresponds to   circular stable orbits, are our only initial assumptions
for the central potential $V(r)$.} 
 The effective potential $U(r)$
 is inverted in each of the two branches
 $r_1(U)\le r_0 $ and $r_2(U)\ge r_0$. 
 }
 \label{fig1}
\end{figure}
\begin{equation}
\label{int}
\int_{r_{\rm min}}^{r_{\rm max}}dr=\int_{E}^{U_{0}}\frac{dr_{1}}{dU}dU+\int_{U_{0}}^{E}\frac{dr_{2}}{dU}dU.
\end{equation}
Performing the change of variables (\ref{int}), we recast equations (\ref{apsi}) and (\ref{te}) into the form 
\begin{equation}
\label{tinv}
\Theta(E,\ell)=\frac{\ell}{\sqrt{2}}\int_{U_{0}}^{E}\frac{1}{\sqrt{E-U}}\frac{d}{dU}\left(\frac{1}{r_{1}}-\frac{1}{r_{2}}\right)dU, 
\end{equation}
and 
\begin{equation}
\label{tenv}
T(E,\ell)=\sqrt{2}\int_{U_{0}}^{E}\frac{1}{\sqrt{E-U}}\frac{d}{dU}(r_{2}-r_{1})dU.
\end{equation}
Equations of the type (\ref{tinv}) and (\ref{tenv}) can be inverted by exploring the well-known Abel integral equation: if $f$ and $g$ are functions such that
\begin{equation}
\label{ap1}
f(u)=\int_{u_{0}}^{u}\frac{g(v)}{\sqrt{u-v}}dv,
\end{equation}
then
\begin{equation}
\label{ap2}
g(v)=\frac{1}{\pi}\frac{d}{dv}\int_{u_{0}}^{v}\frac{f(u)}{\sqrt{v-u}}du.
\end{equation}
For further details and references on the  Abel integral equation, as well as some previous applications, see \cite{AHC,iso1,iso2,iso3,iso4,iso5,Farina,iso6}.
 {It is very important to recall the conditions assuring that (\ref{ap2}) is indeed a solution
of (\ref{ap1}), which can be found, for instance, in   Section 3 of      B\^ocher's  classic  book \cite{bocher}. For our purposes here, it suffices that $f(u)$ be continuously differentiable in 
the interval $I=[u_0,u_1]$ to assure that $g(v)$, not necessarily continuous,  given by (\ref{ap2}) is the unique
solution of (\ref{ap1}) in $I$.}
 {Under the continuous  differentiability  condition, we can  write}  the integrals  (\ref{tinv}) and
(\ref{tenv}), respectively, as}
\begin{eqnarray}
\label{invd}
\frac{1}{r_{1}}-\frac{1}{r_{2}}&=&\frac{\sqrt{2}}{\pi \ell}\int_{U_{0}}^{U}\frac{\Theta(E,\ell)}{\sqrt{U-E}}dE,\\
\label{invt}
r_{2}-r_{1}&=&\frac{1}{\sqrt{2}\pi}\int_{U_{0}}^{U}\frac{T(E,\ell)}{\sqrt{U-E}}dE.
\end{eqnarray}
{The integration constants arising from the use of  (\ref{ap1}) and (\ref{ap2}) 
vanish for both cases, since   we have $r_2(U_0) = r_1(U_0)$, see Fig. \ref{fig1}, and both
integrals in the right-hand sides of (\ref{invd}) and
(\ref{invt}) vanish for $U\to U_0$. 
 }
Notice that equation (\ref{invd}) has already appeared in \cite{YT} in an alternative derivation  of Bertrand's theorem,  whereas equation (\ref{invt}) is already presented, for instance, in Landau and Lifshitz's   textbook \cite{LL}.

\section{Isochrone solutions}
\label{sec3}

The isochrony condition, {\em i.e.}, the requirement that the radial period $T$ does not depend on the angular momentum $\ell$, is fully equivalent to the condition that the {azimuthal} $\Theta$ angle does not depend on the energy $E$. This can be seen from the identity 
\begin{equation}
\label{arp}
 \frac{\partial{T}}{\partial{\ell}}=-2\frac{\partial{\Theta}}{\partial{E}}  ,
\end{equation}
which can be proved directly evoking  the so-called radial action
\begin{equation}
\mathcal{A}_{r}(E,\ell) = \sqrt{2}\int_{r_{\rm min}}^{r_{\rm max}} \sqrt{E-U(r,\ell)}dr,
\end{equation}
and reminding that 
\begin{equation}
\label{idar}
T = 2\frac{\partial \mathcal{A}_{r}}{\partial E}, \quad
\Theta = -\frac{\partial \mathcal{A}_{r}}{\partial \ell}.
\end{equation}
{Hence, 
assuming the isochrony condition
$
\Theta= \pi\lambda(\ell),
$
 {the continuous differentiability condition for the Abel inversion is trivially satisfied and}
equation (\ref{invd}) can be integrated directly as
\begin{equation}
\label{invr}
\frac{1}{r_{1}}-\frac{1}{r_{2}} =  {\beta_{\ell}}\sqrt{U-U_{0}},
\end{equation}
where $\beta_{\ell}=2\sqrt{2}\lambda(\ell)/\ell$. For potentials admitting closed orbits for all values of $\ell$, {\em i.e.}, for the case of Bertrand's theorem, $\lambda$ is a rational number independent of  $\ell$. }  {The left-hand side of equation (\ref{invr}) 
is a smooth ($C^\infty$) function for $U>U_0$,  and this implies that both $r_1(U)$ and $r_2(U)$ are   $C^\infty$  and that, consequently, $U(r)$ will be also a
smooth function, with the only possible exception at its minimum $U_0=U(r_0)$. This can be proved recalling
Fig. \ref{fig1}. Suppose $V(r)$ is not smooth, for instance, at point $r=r_2$. Unless $V(r)$ has an exactly canceling non-smoothness at $r=r_1$, equation (\ref{invr}) will not hold, with the only exception corresponding to the
non-smoothness precisely located at $r=r_0$,   the only common point for the two branches. But even if we admit   a ``fine-tuned'' potential $V(r)$ having exactly canceling non-smooth terms at $r=r_1$ and $r=r_2$, such terms will necessarily depend on $\ell$, and this is excluded by the decomposition of the effective potential (\ref{eff}) in a centrifugal barrier and a pure central potential, which should never depend on $\ell$.
 In summary, we start
with a $C^1$ potential function $U(r)$ and conclude, solely from the isochrony condition, that
$U(r)$ must be $C^\infty$ everywhere, except possibly at $r=r_0$. We will return to this possible
non-smoothness of $U$ at $r_0$ in the last section. 
}

The case of the integral (\ref{invt}) is different. 
 {The isochrony condition does not impose any restriction on the function $T=T(E)$, it is, in principle, a completely arbitrary function}. However, we can write the right-hand  side of (\ref{invt}) in a convenient functional form, without loss of generality, as
\begin{equation}
\label{clo}
r_{2}-r_{1} = \frac{\sqrt{U-U_{0}}}{  h(U,U_0)}, 
\end{equation}
where $ h(U,U_0)$ is an undetermined arbitrary function. {For the sake of notation, we will denote this function simply as $h(U)$.}
{It is important to stress that (\ref{clo}) is merely a definition for the function $h(U)$, there is absolutely no loss of generality in this choice, whose main motivation  comes} from the fact that $U$ must have a local minimum at $r=r_{0}$ in order to guarantee the existence of bounded orbits, see
Fig. \ref{fig1}. A simple Taylor series expansion of $U$  gives us $U-U_{0}\rightarrow [U''(r_{0})/2](r_{1,2}-r_{0})^2$ as $r_{1,2}\rightarrow r_{0}$, so that (\ref{clo}) can be locally verified, with $h(U_{0})=\sqrt{U''(r_{0})/8}$. Thus, (\ref{clo}) captures the essence of behavior of $U(r)$, which must have a local minimum at $r_0$.
{It is important to stress that the Taylor series argument is only a motivation for 
getting (\ref{clo}), no {extra regularity assumption on $U$ around its minimum at
$r_0$ is necessary here. Notice also that, since the left-hand side of (\ref{clo}) is smooth
for $U>U_0$ as a consequence of the isochrony condition, we have that $h(U)$ is also a smooth
function for $U>U_0$.
}
The convenience of the choice (\ref{clo})    will become clear by solving the equations (\ref{invr}) and (\ref{clo}) for the two branches  $r_{1,2}(U)$, leading to
\begin{equation}
\label{2b}
\sqrt{U-U_0}=r_{2}h(U)-\frac{1}{\beta_{\ell}r_{2}}=-\left[r_{1}h(U)-\frac{1}{\beta_{\ell}r_{1}}\right],
\end{equation}
from where we have
\begin{equation}
\label{vit}
U-U_{0}=\left[rh(U)-\frac{1}{\beta_{\ell}r}\right]^{2}.
\end{equation}
As we can see, the choice (\ref{clo}) allows us to obtain symmetrical expressions in (\ref{2b}) for both branches $r_{1,2}(U)$ and, consequently, a unique expression (\ref{vit}) 
valid for all $r$.

The isochrony condition is now equivalent to the existence of solutions of (\ref{vit}) for the effective central potential $U$ {and, rather surprisingly, this is sufficient to constrain
the unknown function $h(U)$!} 
 {The key observation here is that the effective 
potential $U(r)$ is not an arbitrary function of $r$ and $\ell$, but it must have the form (\ref{eff}), with a $V(r)$ which
does not depend on $\ell$. Equation (\ref{vit})  reads
\begin{equation}
\label{ee3}
r^2h^2(U) - \frac{2h(U)}{\beta_\ell} =
V(r) + C(\ell) r^{-2} -U_0  ,
\end{equation} 
with $C(\ell)=\left(\frac{\ell^2}{2} -  \frac{1}{\beta_\ell^2 }\right)$. Notice that $U_0$ can
also depend on $\ell$. 
Equation (\ref{ee3}) fixes the $\ell$-dependent terms of the left-hand  side, and 
this turns out to be a strong restriction on the possible functions $h(U)$. 
For instance, it is easy
to check that (\ref{ee3}) will have solutions of the form (\ref{eff}) for polynomial 
$h(U)$ only for the linear (affine, being more precise) case, {\em i.e.}, only for $h=\alpha U+\gamma$. 
Recall that the effective potential for an attractive $V(r)$ is dominated by the centrifugal
barrier for $r\to 0$. Examining this limit in (\ref{ee3}), from the $r^{2}h^{2}$ term on the left-hand  side, we have that
a polynomial function $h(U)$ of degree $n$, for instance, will give origin to a term proportional to
$\ell^{4n}/r^{4n-2}$, which will be unbalanced with respect  to the right-hand side, unless   $n=1$. 
We will return to the case of more general   $h(U)$ in the last section. 
}  Assuming a linear $h=\alpha U+\gamma$ and multiplying both sides of (\ref{vit}) by $r^2$, we get
\begin{equation}
\label{parabola1}
 \left[\alpha r^2V + \gamma r^2 +\frac{\alpha\ell^2}{2}-\frac{1}{\beta_{\ell}}\right]^{2} -r^2 V -\frac{\ell^2}{2} + r^2U_0 = 0,
\end{equation}
where (\ref{eff}) was used. It is clear that we have re-obtained the Hénon parabolae (\ref{parabola}), with no other assumption than a  {continuously differentiable $V(r)$ and} the isochrony condition.  
The general linear expression $h=\alpha U+\gamma$ reduces to two qualitatively distinct cases: $\alpha=0$ or $\gamma=0$. If both coefficients are non vanishing, we can rewrite $h=\alpha(U+\gamma/\alpha)$, and then $\gamma/\alpha$ can be disregarded, without loss of generality, otherwise, it would mean adding a constant to the potential $V$.

The simplest case here is the constant $h(U)=\sqrt{k/2}$, which gives us the isotropic harmonic potential $V_{\rm ha}(r)$ in (\ref{pot}), together with the closed orbit condition $\lambda=1/2$, and therefore it is a Bertrand's solution. Notice that, in this case, from (\ref{vit}), we also have  $U_0=\ell\sqrt{k}$, as expected for the harmonic potential. 
The second case, $h(U)=\alpha U$, is a  bit more involved. 
We have from (\ref{parabola1}) in this case 
\begin{equation}
\label{iso1}
r^{2}V^{2}-\left(\frac{1}{\alpha^2}-\ell^2+\frac{\ell}{\sqrt{2}\alpha\lambda}\right)V+\left(\frac{U_{0}}{\alpha^2}+\frac{c}{r^2}\right)=0,
\end{equation}
where we have set
\begin{equation}
\label{iso2}
-\frac{\ell^2}{2}\left(\frac{1}{\alpha^2}-\frac{\ell^2}{2}+\frac{\ell}{\sqrt{2}\alpha\lambda}-\frac{1}{4\alpha^{2}\lambda^{2}}\right)=c.
\end{equation}
The first group of solutions for isochrone potentials comes from $c=0$, after introducing the parameters
\begin{equation}
\label{set1}
\frac{1}{\alpha^{2}}-\ell^2+\frac{\ell}{\sqrt{2}\alpha\lambda}=2bk,\qquad -\frac{U_{0}}{\alpha^2}=\pm k^2,
\end{equation}
where $k>0$ and $b\geq0$, resulting in the following attractive potentials
\begin{equation}
\label{vhe}
V_{\mp}(r)=\frac{\mp k}{b+\sqrt{b^{2}\pm r^{2}}},
\end{equation}
which correspond  to the H\'enon and to the bounded potentials $V_{\rm He}(r)$ and $V_{\rm bo}(r)$, respectively, see (\ref{he}) and (\ref{boho}). The parameters $\alpha$ and $\lambda$ leading to these solutions are, respectively
\begin{eqnarray}
\label{ahe}
\alpha&=&\mp\frac{1}{\ell}\frac{1}{\sqrt{1+2bk/\ell^2\pm\sqrt{1+4bk/\ell^{2}}}},\,\,\,\,\,\,\,\,\\
\label{lhe}
\lambda&=&\frac{1}{2}\left(1\pm\frac{1}{\sqrt{1+4bk/\ell^2}}\right).
\end{eqnarray}

The second type of solution arises from setting the parameters
\begin{equation}
\label{set0}
\frac{1}{\alpha^2}-\ell^2+\frac{\ell}{\sqrt{2}\alpha\lambda}=0,
\end{equation}
\begin{equation}
\label{set2}
c=b^{2}k^{2},\qquad -\frac{U_{0}}{\alpha^2}=k^2,
\end{equation}
where $k>0$ and $b\geq0$, resulting in the attractive potential
\begin{equation}
\label{vhe1}
V(r)=-k\frac{\sqrt{r^2-b^2}}{r^2},
\end{equation}
which is the remaining isochrone potential $V_{\rm ho}(r)$, see (\ref{boho}). The parameters $\alpha$ and $\lambda$ for this case are
\begin{eqnarray}
\label{ahe1}
\alpha&=&-\frac{1}{\ell}\frac{1}{\sqrt{1+\sqrt{1+(2bk/\ell^2)^2}}},\\
\label{lhe1}
\lambda&=&\frac{1}{\sqrt{2}}\sqrt{\frac{1+\sqrt{1+(2bk/\ell^2)^2}}{1+(2bk/\ell^{2})^2}}.
\end{eqnarray}
Evidently, the Newtonian potential is a particular case of the H\'enon potential in (\ref{vhe}) and the hollowed potential (\ref{vhe1}), in both case with $b=0$, with the closed orbit condition $\lambda=1$, as also expected from the Bertrand's theorem.
Notice, however, that the H\'enon  (\ref{he}) and the bounded (\ref{boho}) potentials can be written also
as
\begin{eqnarray}
V_{\rm He}(r) &=&  \frac{kb}{r^2}-\frac{k}{r^2}\sqrt{b^2 + r^2},  \\
V_{\rm bo}(r) &=&  \frac{kb}{r^2}-\frac{k}{r^2}\sqrt{b^2 - r^2},    
\end{eqnarray}
from where we see that both potentials $V_\mp(r)$ arising for  $b  <0$ in (\ref{set1}) correspond to some
 $(\varepsilon,\Lambda)$-gauge redefinitions of standard isochrone potentials, completing all solutions
 we can obtain from (\ref{parabola1}).
}

\section{Isochrone potentials are Keplerian}
\label{sec4}

Our approach allows the exact determination of the radial period for all isochrone potentials. Notice that equations  {(\ref{tenv})} and (\ref{clo}) imply that
 {
\begin{equation}
\label{tenvv}
T(E)=\sqrt{2}\int_{U_{0}}^{E}\frac{1}{\sqrt{E-U}}\frac{d}{dU}\left(\frac{\sqrt{U-U_{0}}}{  h(U)}
\right)
dU.
\end{equation}
}
For the isotropic harmonic potential case, we know from the last section that $h(U)=\sqrt{k/2}$, and therefore
\begin{equation}
\label{tha}
T(E)=\frac{\pi}{\sqrt{k}}.
\end{equation}
The other group of isochrone potentials are more interesting. By using   $h(U)=\alpha U$ and   $|U_{0}|=\alpha^{2}k^{2}$, Eq. (\ref{tenvv}) yields
\begin{equation}
\label{tkep}
T(E)=\frac{\pi k}{\sqrt{2|E|^{3}}},
\end{equation}
which is exactly a Keplerian equation, independent of $b$. Moreover, invoking the
Abel inversion, the radial period (\ref{tkep}) implies  $h=\alpha U$ with $|U_{0}|=\alpha^{2}k^{2}$.

We can also obtain the Kepler's third law 
\begin{equation}
\label{kep3}
T^2 = \frac{4\pi^2}{k}a^3
\end{equation}
 for these models, see \cite{PPP} for the usual derivation of this result. In order have a
Kepler's third law, we have to find an orbital characteristic length $a$ which is inversely proportional to the energy $E$. The starting point is the  periapsis and apoapsis equation, {\em i.e.}, $V(r)+\ell^{2}/2r^{2}=E$. For the isochrone potentials (\ref{vhe}) we define $\xi_{\pm}=\sqrt{b^{2}\pm r^{2}}$, which   satis\-fies
\begin{equation}
\label{root1}
|E|\xi_{\pm}^{2}-k\xi_{\pm}+\left(kb+\frac{\ell^2}{2}-|E|b^{2}\right)=0,
\end{equation}
and therefore
\begin{equation}
\label{d1}
a_{\pm}=\frac{\sqrt{b^{2}\pm r_{\rm max}^{2}}+\sqrt{b^{2}\pm r_{\rm  min}^{2}}}{2}=\frac{k}{2|E|}.
\end{equation}
Similarly, for the isochrone potential (\ref{vhe1}) we now define $\xi=\sqrt{r^{2}-b^{2}}$, which satisfies
\begin{equation}
\label{root2}
|E|\xi^{2}-k\xi+\left(\frac{\ell^2}{2}-|E|b^{2}\right)=0,
\end{equation}
and therefore
\begin{equation}
\label{d2}
a=\frac{\sqrt{r_{\rm max}^{2}-b^{2}}+\sqrt{r_{\rm min}^{2}-b^{2}}}{2}=\frac{k}{2|E|}.
\end{equation}
For  the cases with $b=0$, $a$ corresponds to the semi-major axis of Kepler's problem.

The fact that the   isochrone potentials (\ref{vhe}) and (\ref{vhe1}) have the 
same Keplerian period (\ref{kep3})  is not a mere coincidence. We can solve the equation of motion (\ref{ene})
\begin{equation}
\label{tr1}
t=\frac{1}{\sqrt{2}}\int\frac{1}{\sqrt{E-V(r)-\ell^{2}/2r^{2}}}dr,
\end{equation} 
for all these potentials
so that it has effectively a Keplerian form 
\begin{equation}
\label{tr2}
t=\frac{1}{\sqrt{2}}\int\frac{1}{\sqrt{E_{*}+k/\xi-\ell_{*}^{2}/2\xi^{2}}}d\xi,
\end{equation}
by performing the change of variable  $\xi_{\pm}=\sqrt{b^{2}\pm r^{2}}$ for the isochrone potentials (\ref{vhe}), with orbital parameters
\begin{equation}
\label{op1}
E_{*}=\pm E,\qquad \frac{\ell_{*}^{2}}{2}=\frac{\ell^{2}}{2}+kb+E_{*}b^{2},
\end{equation}
and the change of variable $\xi=\sqrt{r^{2}-b^{2}}$ for the isochrone potential (\ref{vhe1}), with orbital parameters
\begin{equation}
\label{op2}
E_{*}=E,\qquad \frac{\ell_{*}^{2}}{2}=\frac{\ell^{2}}{2}-E_{*}b^{2}.
\end{equation}
Setting the parameters  
\begin{equation}
\label{pec}
p=\frac{\ell_{*}^{2}}{k},\qquad e=\sqrt{1+\frac{2E_{*}\ell_{*}^{2}}{k^{2}}},
\end{equation} 
we have an effective  
Kepler's problem
leading to the following parametric solutions 
\begin{equation}
\label{eme}
\xi=a(1-e\cos\psi),\qquad t=\sqrt{\frac{a^3}{k}}(\psi-e\sin\psi),
\end{equation}
\begin{equation}
\label{eze}
\xi=\frac{p}{2}(1+\psi^2),\qquad t=\frac{1}{2}\sqrt{\frac{p^3}{k}}\left(\psi+\frac{\psi^3}{3}\right),
\end{equation}
\begin{equation}
\label{ema}
\xi=a(e\cosh\psi-1),\qquad t=\sqrt{\frac{a^3}{k}}(e\sinh\psi-\psi),
\end{equation}
for $E_{*}<0$, $E_{*}=0$, and $E_{*}>0$, respectively, where we use the initial condition that $\xi$ takes its smallest value at $t=0$ for all cases. The corresponding values of $r$ can be simply redeemed by reversing the respective $\xi$-transformation.

\section{Conclusion}
\label{sec5}

 {We have obtained the Hénon’s isochrone potentials by exploring the Abel 
integral inversion, under mild smoothness assumptions on the potential function. In sharp contrast with the
usual derivations in the literature, which typically require analyticity of the potential function $V(r)$, 
our approach only demands a $C^1$ function. Two points of our analysis deserve further comments. 
First, we have shown that the isochrony condition implies a smooth $C^\infty$ potential function $V(r)$
everywhere, except possibly at the minimum of the effective potential at $r = r_0.$ This could suggest it would be worth seeking
 for isochrone potentials corresponding to the $C^1$ matching at $r_0$ of $C^\infty$ functions. However, since
 $r_0$ does depend on $\ell$, the final $C^1$ function would also depend on the angular momentum, and hence
 it would correspond to a nonphysical potential. The second point we wish to stress is related to the
  non-polynomial    $h(U)$ cases. Essentially, the argument 
we have used in Sect. III rules out any function $h(U)$ which does not have a linear asymptotics for
large $U$. However, this is not a strong restriction among non-polynomial functions. We could consider, 
for instance, any rational function $h(U) = P_{n+1}(U)/Q_n(U)$.
In this case,  we can show   that, for $n>0$,  equation (\ref{ee3}) will be a higher-order polynomial in $U$, whose roots will hardly have the required form (\ref{eff}). 
The existence of such roots for any other case besides the linear (affine) $h(U)$ case would imply the existence of new families of isochrone potentials, something
quite improbable. Unfortunately, we could not prove rigorously that only the linear case
gives origin to acceptable solutions with the required form (\ref{eff}). Notwithstanding, this does
not compromise our basic result: the derivation of   Henon's isochrone potentials by using the
Abel inversion theorem 
 with minimal regularity
assumptions on the potential function $V(r)$.
 }

 {As we have said previously, our approach is quite general
and can be used in other contexts as well.}
The isochrone potentials have arisen from the condition $T(E,\ell) = T(E)$, which implies $\Theta(E,\ell) = \Theta(\ell)$. We can easily determine, for instance, the potentials arising from the condition $T(E,\ell) = T(\ell)$. Notice that from (\ref{idar}), we have that $\partial _E T = 0$ implies that $\mathcal{A}_r$ is  a linear function in $E$, which for sake of convenience we write as
\begin{equation}
\mathcal{A}_r = \frac{\pi}{2\sqrt{2}}\left(  Eg(\ell) - f(\ell)\right),
\end{equation}
which leads to
\begin{equation}
\label{aa}
T =  \frac{\pi}{ \sqrt{2}}g(\ell), \quad \Theta = \frac{\pi}{2\sqrt{2}}(f'(\ell) - Eg'(\ell)).
\end{equation}
 {Since both $T$ and $\Theta$ are smooth functions in $E$, we can  invoke 
the Abel integral inversion for both functions.  Integrating} (\ref{invd}) and (\ref{invt}) we obtain
\begin{eqnarray}
\label{invd1} 
\frac{1}{r_{1}}-\frac{1}{r_{2}}&=&{ \frac{1}{\ell} \left(f'(\ell) - (2U+U_0)\frac{g'(\ell)}{3}\right)\sqrt{U-U_0}}
 ,\\
\label{invt1}
r_{2}-r_{1}&=& g(\ell)  \sqrt{U-U_0}.
\end{eqnarray}
Proceeding as in the isochrone problem, we get, instead of (\ref{vit}), the following equation which is valid for the two branches $r_1(U)$ and $r_2(U)$
 \begin{equation}
\label{vit1}
 {
U-U_{0}=\left[\frac{r}{g(\ell)} - \frac{\ell}{ r\left(f'(\ell) - (2U+U_0)\frac{g'(\ell)}{3}\right)}\right]^{2}.}
\end{equation}
The solutions of (\ref{vit1}) of the form (\ref{eff}) are the potentials with radial period independent of $E$. Essentially, we have two distinct cases. The first one corresponds to $g'=0$. In this case, $T$ is also constant and the potential is an isochrone one.  It is easy to see from (\ref{vit1}) that this is the case of the harmonic potential 
 with a possible extra (gauge) term $\Lambda/r^2$, corresponding to 
 {
\begin{equation}
f'(\ell)^2 = \frac{2\ell^2}{2\Lambda + \ell^2}.
\end{equation}
The case with $g'(\ell)\ne 0$ is much more involved and (\ref{vit1}) boils down to a cubic equation for $U$, with no solutions of the type (\ref{eff}).
In summary, the condition $T(E,\ell) = T(\ell)$ is sufficient to select isochrone potentials of the harmonic type.}

We finish by noticing that our results  
  imply       Bertrand's theorem.
From (\ref{arp}), we have that any potential satisfying Bertrand's theorem must be isochrone. By inspecting $\lambda$ determined in the Section \ref{sec3}, we conclude immediately that only the Newtonian and the harmonic potentials have {azimuthal} angle $\Theta$ independent of $E$ and $\ell$.

\begin{acknowledgements}
AS acknowledges the financial support of
CNPq and FAPESP (Brazil) through the grants, respectively,  302674/2018-7 
and  21/09293-7. 

\end{acknowledgements}


\begin{thebibliography}{99}

 
\bibitem{MH1} M. H\'enon, ``L'amas Isochrone I'', Annales d'Astrophysique {\bf 22}, 126 (1959).

\bibitem{MH2} M. H\'enon, ``L'amas Isochrone II'', Annales d'Astrophysique {\bf 22}, 491 (1959).


\bibitem{MH3} M. H\'enon, ``L'amas Isochrone III'',  Annales d'Astrophysique {\bf 23}, 474 (1960).

\bibitem{JBY} J. Binney, ``H\'enon's Isochrone Model'', in {\it Une vie d\'edi\'ee aux syst\`emes dynamiques: Hommage \`a Michel H\'enon} ed. J.-M. Alimi, R. Mohayaee \& J. Perez, Hermann, 2016, p. 99-109 - arXiv:1411.4937.

\bibitem{BTGD} J. Binney and S. Tremaine, {\it Galactic Dynamics}, 2nd ed. (Princeton University Press, Princeton, 2008).

\bibitem{PPD} A. Simon-Petit, J. Perez, and G. Duval, ``Isochrony in 3D Radial Potentials. From Michel H\'enon's Ideas to Isochrone Relativity: Classification, Interpretation and Applications'', Commun. Math. Phys. {\bf 363}, 605 (2018).

\bibitem{PPP} A. Simon-Petit, J. Perez, and G. Plum, ``The status of isochrony in the formation and evolution of self-gravitating systems'', Mon. Not. R. Astron. Soc. {\bf 484}, 4963-4971 (2019).

\bibitem{PPC} P. Ramond and J. Perez, ``The Geometry of Isochrone Orbits: from Archimedes' parabolae to Kepler's third law'', Cel. Mech. \& Dyn. Astro {\bf 132}, 22 (2020).

\bibitem{RP} 
P. Ramond and J. Perez, {``New Methods of Isochrone Mechanics''},  {	J. Math. Phys. {\bf 62}, 112704 (2021).} [arXiv:2104.05643]
 
\bibitem{R1} K.P. Rauch, S. Tremaine, {\em  Resonant Relaxation in Stellar Systems},
New Astronomy {\bf 1},  149 (1996).
[arXiv:astro-ph/9603018]



\bibitem{R2}M. Atakan Gurkan  and  C. Hopman, {``Resonant relaxation near a massive black hole: the dependence on eccentricity''}, Mon. Not. Roy. Astron. Soc. {\bf 379}, 1083 (2007). [arXiv:0704.2709]


\bibitem{R3}Y. Meiron and B. Kocsis, {``Resonant relaxation in globular clusters''}, 
 Astrophysical J. {\bf 878},  138 (2019).
[arXiv:1806.07894]  

\bibitem{Abel} N.H. Abel, “Aufl\"osung einer mechanischen Aufgabe”, J. Reine Angew.
Math. {\bf 1}, 153  (1826).

\bibitem{LL} L.D. Landau and E.M. Lifshitz, {\it Mechanics} (Pergamon Press, London, 1982).

\bibitem{AHC} A. H. Carter, ``A class of inverse problems in physics'',  Am. J. Phys. {\bf 68}, 698-703 (2000).


\bibitem{iso1} E.T. Osypowski and M. G. Olsson, “Isynchronous motion in classical me-
chanics”, Amer. J. Phys. {\bf 55}, 720  (1987).


\bibitem{iso2} O.A. Chalykh and A.P. Veselov, “A remark on rational isochronous potentials”, J. Nonl. Math. Phys. {\bf 12}, Suppl, 179  (2005). [arXiv:math-ph/0409062]

\bibitem{iso3} J. Dorignac, “On the quantum spectrum of isochronous potentials”, J.
Phys. A {\bf 38}, 6183  (2005). [arXiv:quant-ph/0504074]


\bibitem{iso4} M. Asorey, J.F. Carinena, G. Marmo and A.M. Perelomov, “Isoperiodic
classical systems and their quantum counterparts”, Ann. Phys. {\bf 322}, 1444
  (2007). [arXiv:0707.4465]

\bibitem{iso5} J.F. Carinena, A.M. Perelomov and M.F. Ranada, “Isochronous classical
systems and quantum systems with equally spaced spectra”, J. Phys. Conf. Ser. {\bf 87}, 012007 (2007).


\bibitem{Farina} {P. Terra, R.M. Souza, C. Farina, ``Is the tautochrone curve unique?'',
Am. J. Phys. {\bf 84}, 917 (2016).
[arXiv:1610.01006]}

\bibitem{iso6} D.J. Cross, “Every isochronous potential is shear-equivalent to a harmonic
potential”, Amer. J. Phys. {\bf 86}, 198  (2018).



\bibitem{bocher}  {M. B\^ocher, {\em An Introduction to the Study of Integral Equations}, Cambridge
University Press, 1909. }



\bibitem{YT} Y. Tikochinsky, {``A simplified proof of Bertrand's theorem''}, Am. J. Phys. {\bf 56}, 1073-1075 (1988).

 
 











\end{thebibliography}
\end{document}